\documentclass{PoS}
\pdfoutput=1

\usepackage[utf8]{inputenc}
\usepackage{amsmath}
\usepackage{amssymb}
\usepackage{graphicx}
\usepackage{xspace}
\numberwithin{equation}{section}
\allowdisplaybreaks[2]
\newcommand{\MINLO}{MiN\protect\scalebox{0.8}{LO}\xspace}
\newcommand{\UNtwoLOPS}{U\protect\scalebox{0.8}{N}\ensuremath{^{2}}L\protect\scalebox{0.8}{O}P\protect\scalebox{0.8}{S}\xspace}

\newcommand{\MCatNLO}{M\protect\scalebox{0.8}{C}@N\protect\scalebox{0.8}{LO}\xspace}

\newcommand{\Powheg}{P\protect\scalebox{0.8}{OWHEG}\xspace}

\newcommand{\MEPS}{M\scalebox{0.8}{E}P\scalebox{0.8}{S}@LO\xspace}
\newcommand{\LOPS}{L\scalebox{0.8}{O}P\scalebox{0.8}{S}\xspace}
\newcommand{\NLOPS}{N\scalebox{0.8}{LO}P\scalebox{0.8}{S}\xspace}
\newcommand{\NNLOPS}{N\scalebox{0.8}{NLO}P\scalebox{0.8}{S}\xspace}
\newcommand{\MENLOPS}{ME\protect\scalebox{0.8}{NLO}PS\xspace}
\newcommand{\MEPSatNLO}{M\scalebox{0.8}{E}P\scalebox{0.8}{S}@N\scalebox{0.8}{LO}\xspace}
\newcommand{\MEPSatLoop}{M\scalebox{0.8}{E}P\scalebox{0.8}{S}@L\scalebox{0.8}{OOP}$^2$\xspace}

\newcommand{\Geneva}{G\protect\scalebox{0.8}{ENEVA}\xspace}
\newcommand{\Herwig}{H\protect\scalebox{0.8}{ERWIG}\xspace}

\newcommand{\Alpgen}{A\protect\scalebox{0.8}{LPGEN}\xspace}

\newcommand{\MGfiveAMC}{M\protect\scalebox{0.8}{AD}G\protect\scalebox{0.8}{RAPH}5\_\aMCatNLO\xspace}

\newcommand{\Pythia}{P\protect\scalebox{0.8}{YTHIA}\xspace}

\newcommand{\aMCatNLO}{aM\protect\scalebox{0.8}{C}@N\protect\scalebox{0.8}{LO}\xspace}
\newcommand{\Sherpa}{S\protect\scalebox{0.8}{HERPA}\xspace}

\newcommand{\done}{{\rm d}}

\newcommand{\GeV}{\,\mathrm{GeV}}

\newcommand{\mr}[1]{\mathrm{#1}}

\newcommand{\pT}{\ensuremath{p_{\mr{T}}}}
\newcommand{\Qcut}{\ensuremath{Q_{\mr{cut}}}}

\title{Parton shower matching and merging}

\ShortTitle{Parton shower matching and merging}

\author{\speaker{Marek Sch\"onherr}\thanks{M.S.\ was supported by PITN--GA--2012--315877 ({\it MCnet}) and the ERC Advanced Grant MC@NNLO (340983).}\\
        Theoretical Physics Department, CERN, Geneva, Switzerland\\
        E-mail: \email{marek.schoenherr@cern.ch}}

\abstract{
  With Run-II of the LHC starting its final year of data-taking, high 
  precision theoretical predictions through Monte-Carlo event generators 
  whose uncertainties match that of the recorded data are of highest 
  importance. 
  This talk summarises the progress of the field and highlight a number 
  of recent developments.
}

\FullConference{Corfu Summer Institute 2017 'School and Workshops on Elementary Particle Physics and Gravity'\\
		2-28 September 2017\\
		Corfu, Greece}

\begin{document}

\section{Introduction}

Modern Monte-Carlo event generators like \Pythia \cite{arXiv:1410.3012},
\Herwig \cite{arXiv:0803.0883,arXiv:1705.06919} or 
\Sherpa \cite{arXiv:0811.4622,hep-ph/0109036,arXiv:0808.3674,
              arXiv:0709.1027,arXiv:1506.05057,hep-ph/0311085,
              arXiv:0810.5071,arXiv:1606.08753}
are the workhorses for the physics analyses and measurements at the LHC. 
In many cases, the \Pythia and \Herwig generators (or their older 
predecessors) receive input from parton level tools which compute 
the hard core production matrix elements either at LO 
(\Alpgen \cite{hep-ph/0206293} or \MGfiveAMC \cite{arXiv:1405.0301}), 
NLO (\MGfiveAMC or \Powheg \cite{arXiv:1002.2581}) 
or even NNLO (\Powheg-\MINLO \cite{arXiv:1309.0017,arXiv:1407.2940,
  arXiv:1603.01620,arXiv:1804.08141} or \Geneva \cite{arXiv:1508.01475}), 
which are matched to the parton shower. 
The following contribution thus highlights a few important improvements 
that become available in recent years, some of the forming the 
standard for the experiments now.

\section{Matching next-to-leading order matrix elements to parton showers}

\begin{figure}[t!]
  \centering
  \includegraphics[width=0.47\textwidth]{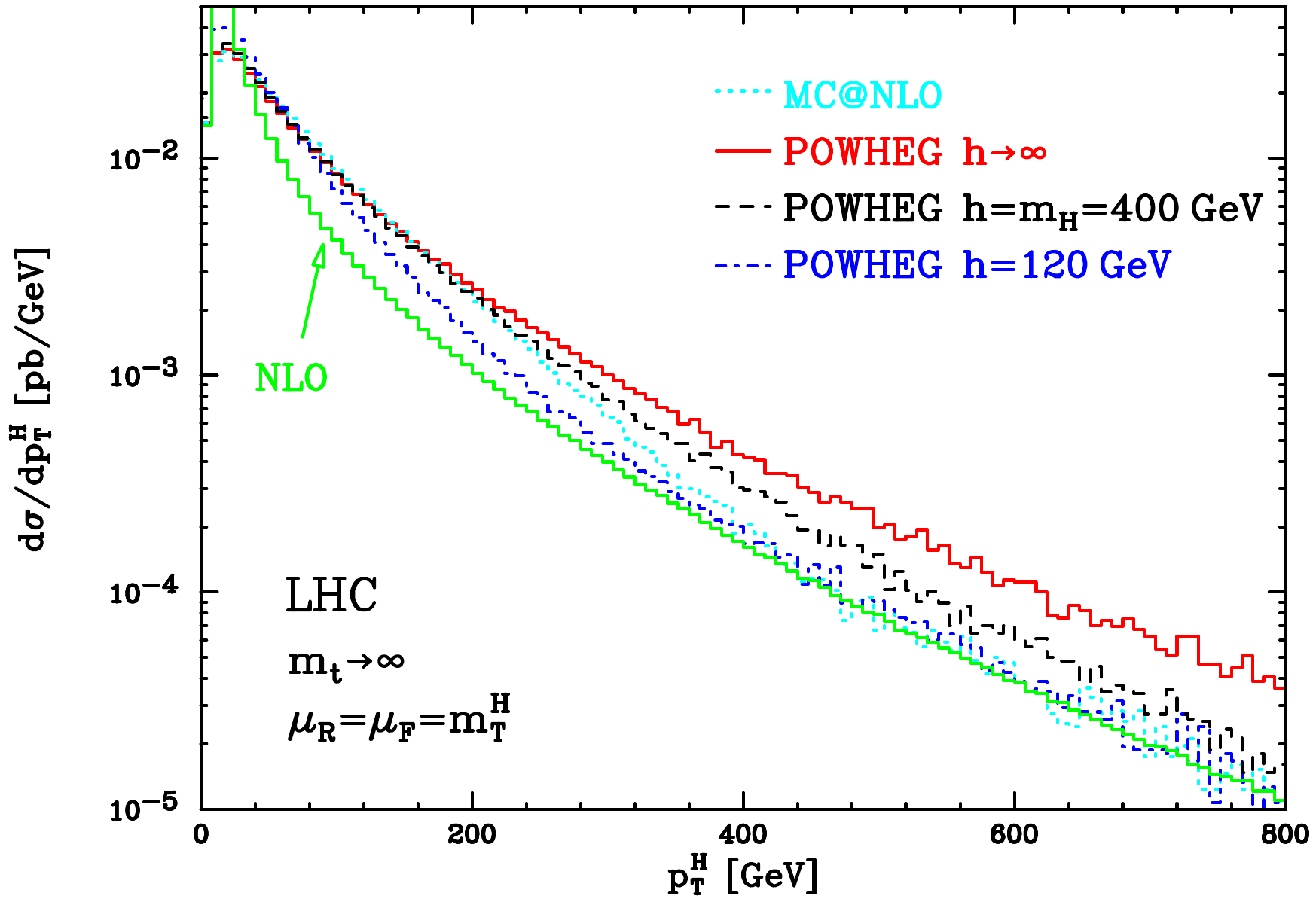}\hfill
  \begin{minipage}{0.44\textwidth}
    \vspace*{-2mm}
    \includegraphics[width=\linewidth]{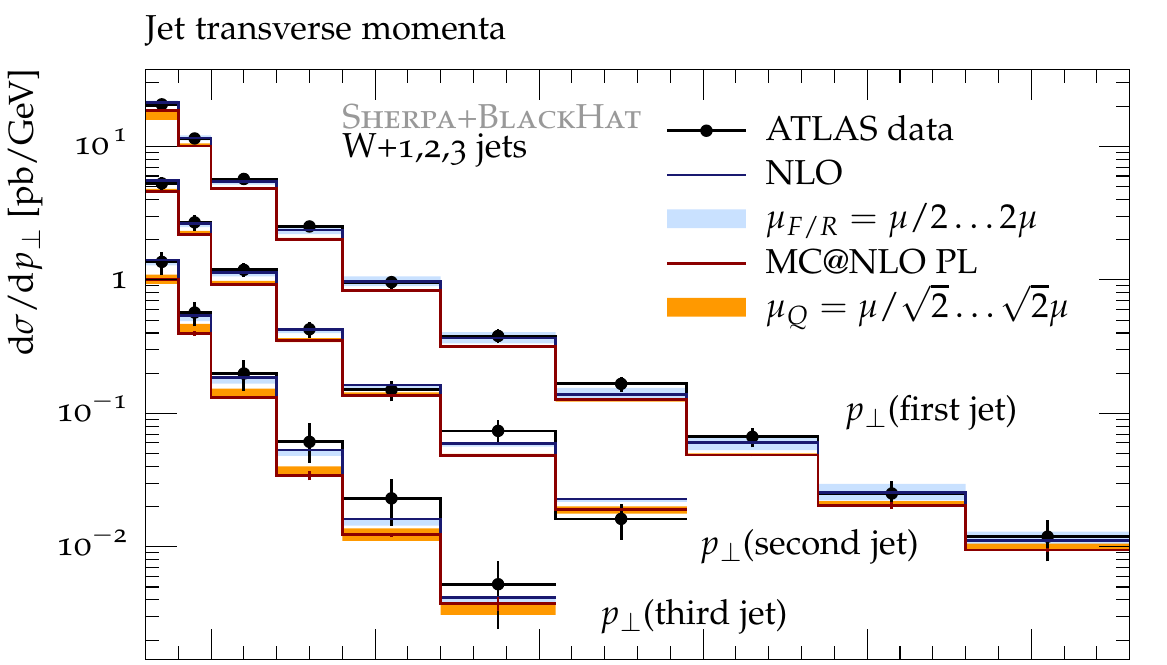}\vspace*{-2.3pt}\\
    \includegraphics[width=\linewidth]{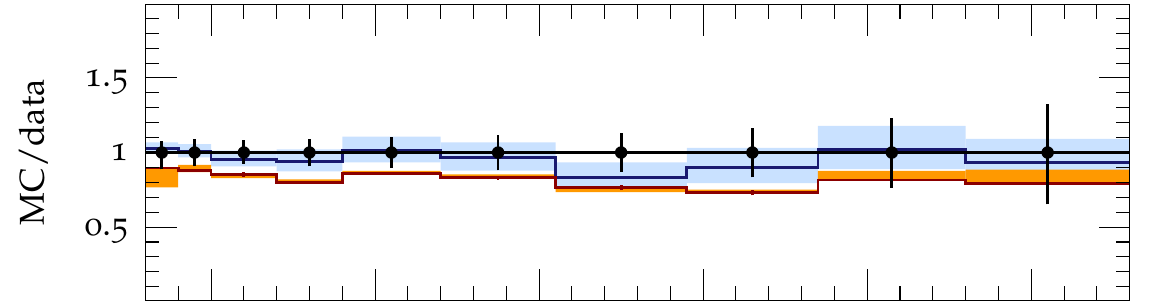}\vspace*{-2.3pt}\\
    \includegraphics[width=\linewidth]{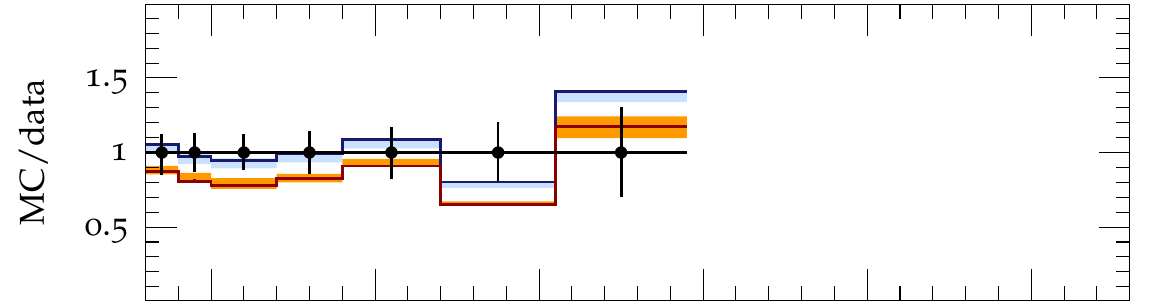}\vspace*{-2.3pt}\\
    \includegraphics[width=\linewidth]{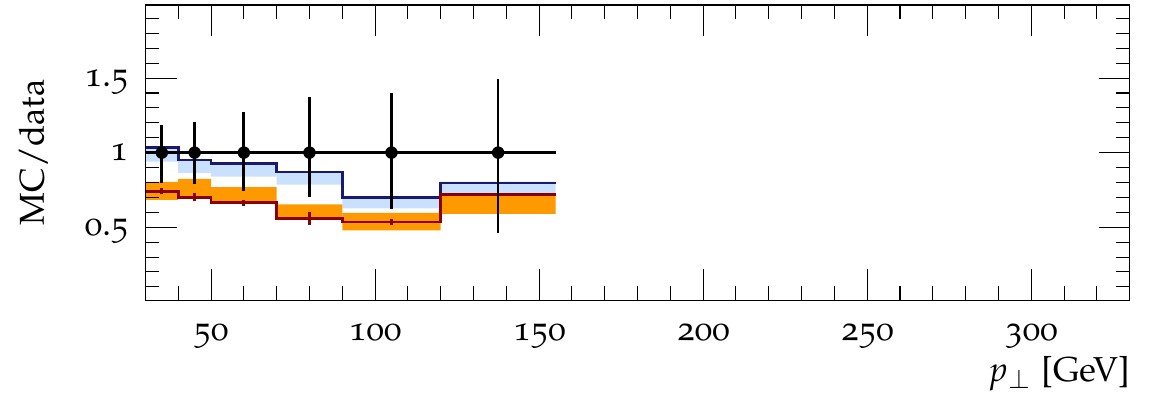}
  \end{minipage}
  \caption{\textbf{Left:} Comparison of the \pT\ spectrum of the Higgs boson ($m_h=400\,\GeV$) 
           in gluon fusion with \MCatNLO and \Powheg, with and without the \texttt{hfact}
           treatment. Figure taken from \cite{arXiv:0812.0578}.
           \textbf{Right:} Parton shower matched results for $W+n$\,jets production 
           using an \MCatNLO-like techniques generated by \Sherpa. 
           Figure taken from \cite{arXiv:1201.5882}.
           \label{fig:powheg-mcatnlo}}
\end{figure}

In a first step, the matching of NLO matrix elementsto parton showers, 
which is known for over a decade now, is briefly reviewed. 
In the literature, there exist various different ansatzes: \Powheg 
\cite{hep-ph/0409146,arXiv:0709.2092} and \MCatNLO 
\cite{hep-ph/0204244,hep-ph/0305252}, and various variants therof 
\cite{arXiv:1109.6256,arXiv:1502.00925,arXiv:1111.1220,arXiv:1201.5882,
      arXiv:1208.2815,arXiv:1309.5912,arXiv:1711.02568,arXiv:1711.03319}.
They are available for all processes of interest and, thus, for the 
minimum baseline standard in LHC physics analyses. 
Their general aim is to keep both the NLO accuracy in an expansion of 
the cross section in $\alpha_s$ at the same time as the full 
logarithmic accuracy of the parton shower resummation. 
\begin{equation}
  \begin{split}
    \langle O\rangle^\text{\NLOPS}
    \,=\;&\int\done\Phi_B\;
	  \overline{\mr{B}}(\Phi_B)\;\widetilde{\mr{PS}}_B(\mu_Q^2,O)
	 +\int\done\Phi_R\;\mr{H}(\Phi_R)\;\mr{PS}_R(t_R,O)
  \end{split}
\end{equation}
with the familiar $\overline{\mr{B}}$ function defined as 
$\overline{\mr{B}}=\mr{B}+\mr{V}+\mr{I}_\mr{K}$, and the hard remainder
$\rm{H}=\mr{R}-\mr{D}_\mr{K}$. 
The matched parton shower $\widetilde{\mr{PS}}_B(\mu_Q^2,O)$ is defined 
through $\widetilde{\mr{PS}}_B(\mu_Q^2,O)=\Delta_{\mr{D}_\mr{K}}(\mu_Q^2,t_c)\,O(\Phi_B)+\int\done t'\,\frac{\mr{D_K}}{\mr{B}}\,\Delta_{\mr{D}_\mr{K}}(\mu^2,t')\,\mr{PS}_R(t',O)$ 
with a continuing standard parton shower $\mr{PS}_n(t,O)$ operating 
on the $n$-parton configuration with starting scale $t$.
The splitting kernels of the matched shower are the $\mr{D}_\mr{K}$ 
and the $\mr{D}_\mr{K}$ are their integrated version.
The resummation region is limited by $\mu_Q$. 
The various approaches now differ in their choices of $\mr{D}_\mr{K}$ and 
$\mu_Q$, they are detailed in Tab.\ \ref{tab:powheg-mcatnlo}. 
While \MCatNLO retains the parton shower's splitting function and 
resummation region definition, \Powheg uses the partitioned real 
emission matrix element as resummation kernel and fills the entire 
available phase space. Both cases can lead to articfacts when large 
NLO corrections are present as, upon expansion of the matched parton 
shower emission, the emission spectrum is enhanced by a factor of 
$\overline{\mr{B}}/\mr{B}$ in the resummation region. 
In \Powheg, these effects can be mitigated using the 
\texttt{hfact}-treatment \cite{arXiv:0812.0578}. 
Its results are shown in Fig.\ \ref{fig:powheg-mcatnlo}.

\begin{table}[b]
  \centering
  \begin{tabular}{l|c|c}
    & \MCatNLO & \Powheg \\\hline
    $\mr{D}_\mr{K}$ & $\rm{B}\cdot\widetilde{\mr{K}}_\mr{PS}$ & $\mr{R}$ \\
    $\mu_Q^2$ & $\mu_F^2$ & $S_\text{had}$
  \end{tabular}\\[2mm]
  \caption{Choices of resummation kernels and the size of the resummation 
           region in \Powheg- and \MCatNLO-type matching algorithms.
           \label{tab:powheg-mcatnlo}}
\end{table}

\begin{figure}[t!]
  \includegraphics[width=0.47\textwidth]{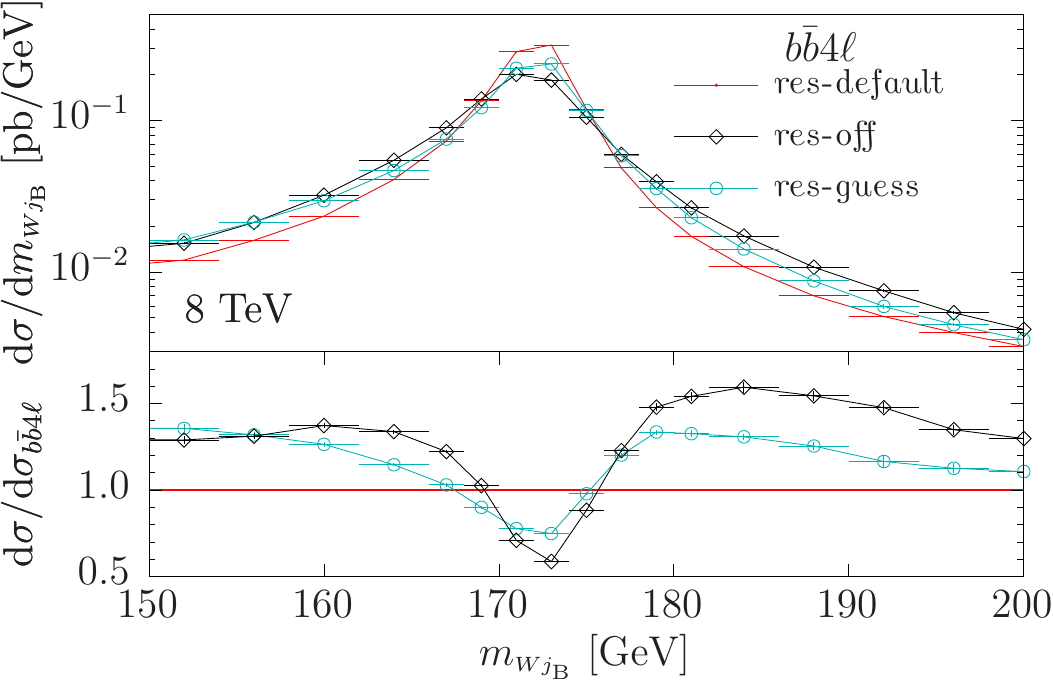}\hfill
  \includegraphics[width=0.47\textwidth]{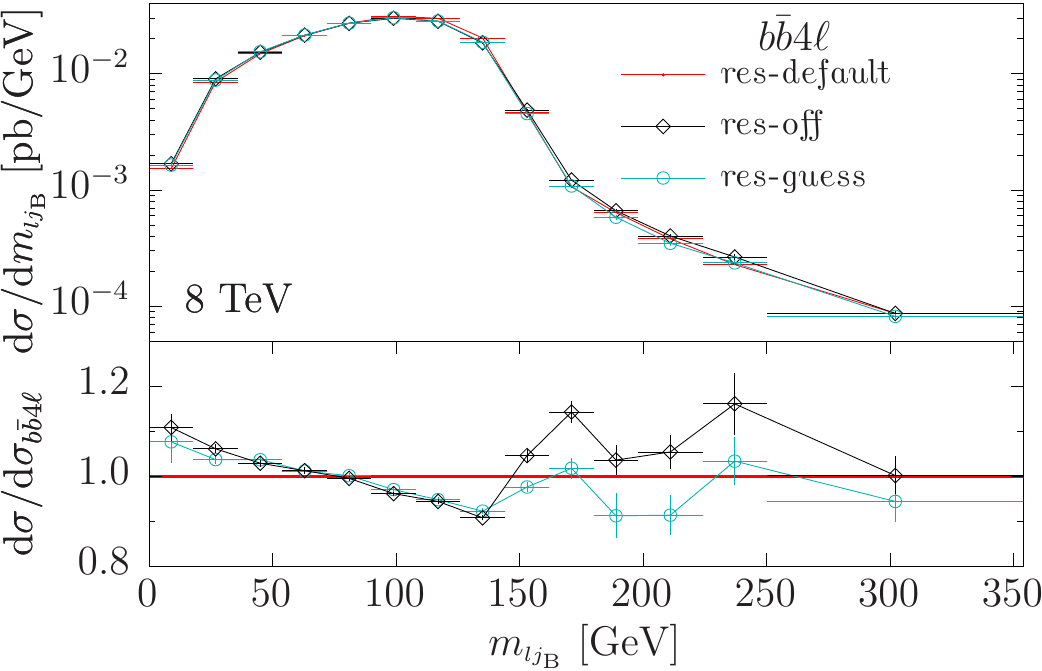}
  \caption{
    Size of the distortion of resonance shape in $t\bar{t}$ production 
    if not treated properly. 
    Figure taken from \cite{arXiv:1607.04538}.
    \label{fig:res-aware}
  }
\end{figure}

A major recent development for processes with internal resonances 
is discussed thereafter has been published in 
\cite{arXiv:1511.08692,arXiv:1603.01178,arXiv:1607.04538,arXiv:1803.00950}. 
Therein, the inherent subtractions and momentum maps have been 
changed such that the matched results preserves the shape of the 
internal resonance and the matching to the parton shower does not 
introduce distortions.
Fig.\ \ref{fig:res-aware} shows that for important observables 
these distortions can amount to more than 50\%, if the matching 
is unaware of the internal resonance.

\begin{figure}[t!]
  \includegraphics[width=0.47\textwidth]{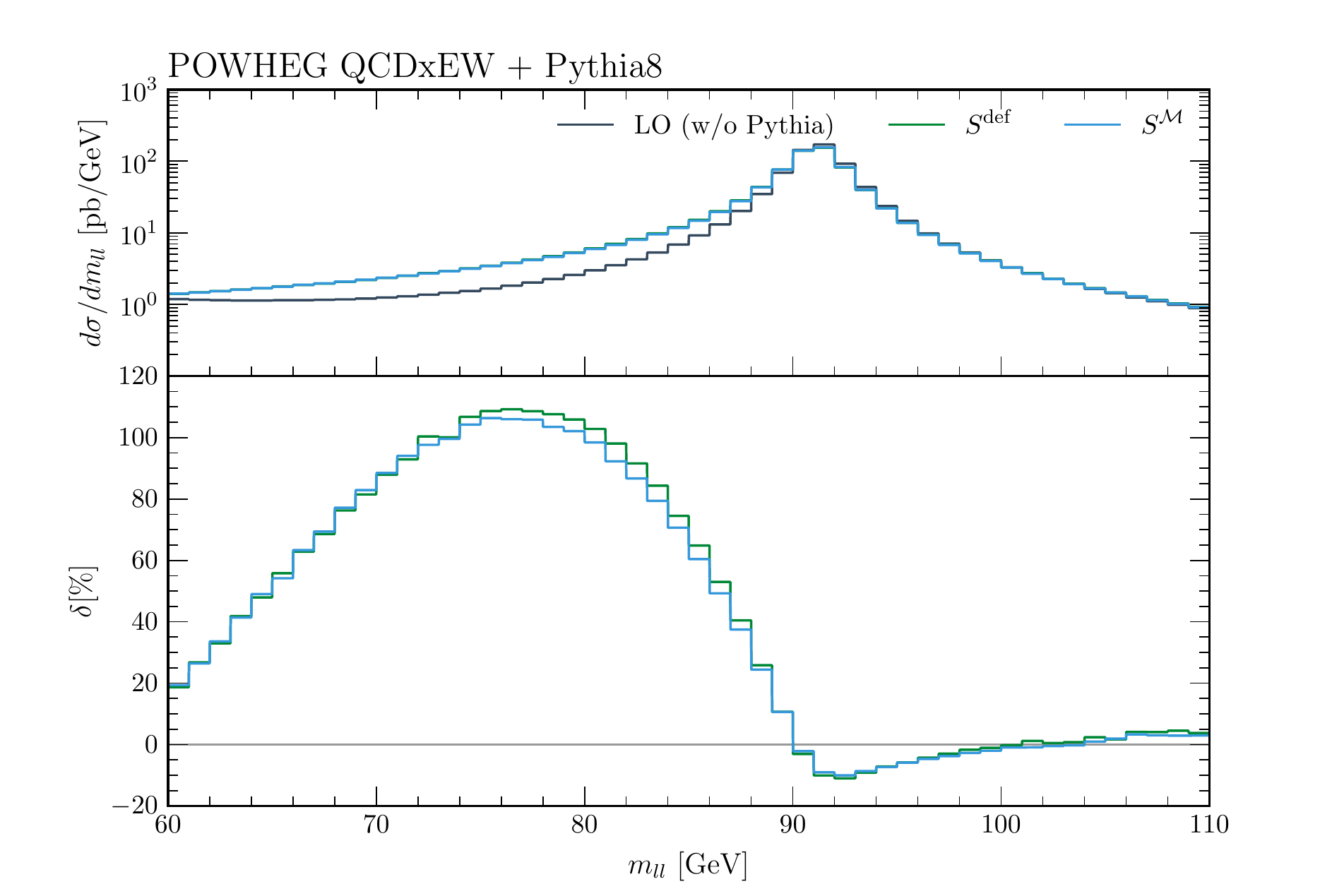}
  \hfill
  \includegraphics[width=0.47\textwidth]{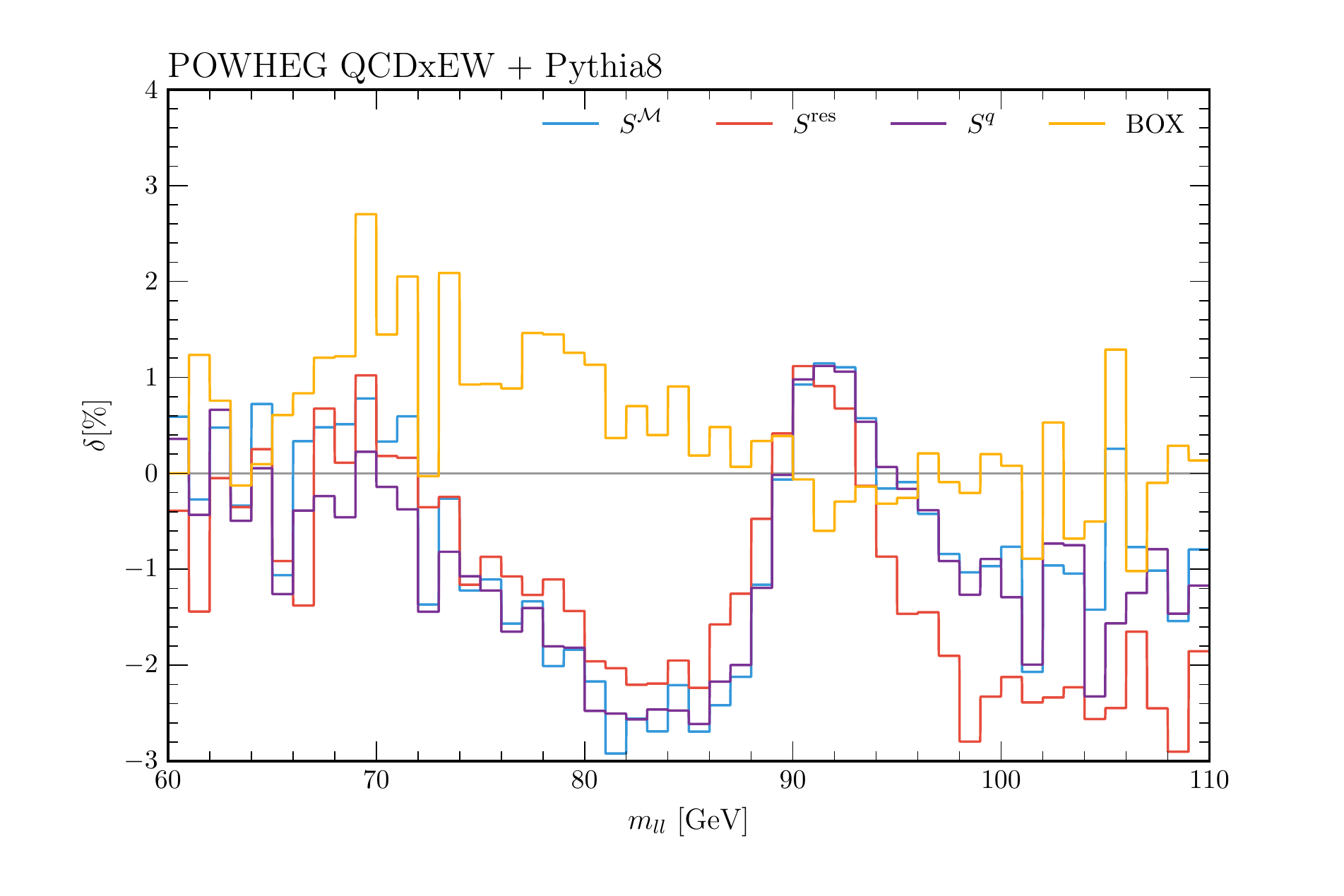}
  \caption{
    NLO QCD+EW parton shower matched calculation for $W$ production. 
    The corrections crucially depend on the resonance awareness of the 
    matching algorithm.
    Figure taken from \cite{arXiv:1612.04292}.
    \label{fig:nlops-ew}
  }
\end{figure}

Finally, a small class of NLO EW corrections has also been matched 
to a QED parton shower \cite{arXiv:1201.4804,arXiv:1302.4606,arXiv:1612.04292}. 
Here, an interleaving with the NLO QCD corrections and a resonance 
aware matching is essential. 
Selected results are shown in Fig.\ \ref{fig:nlops-ew}.

\section{Matching next-to-next-to-leading order matrix elements to parton showers}

\begin{figure}[t!]
  \includegraphics[width=0.47\textwidth]{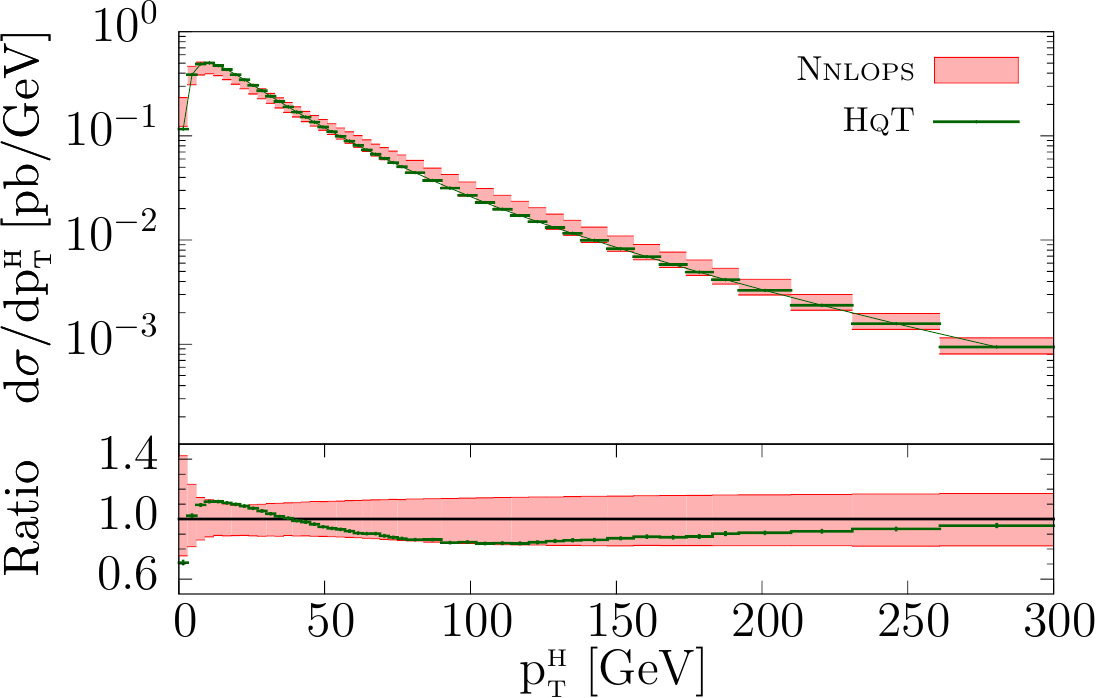}\hfill
    \includegraphics[width=0.47\textwidth]{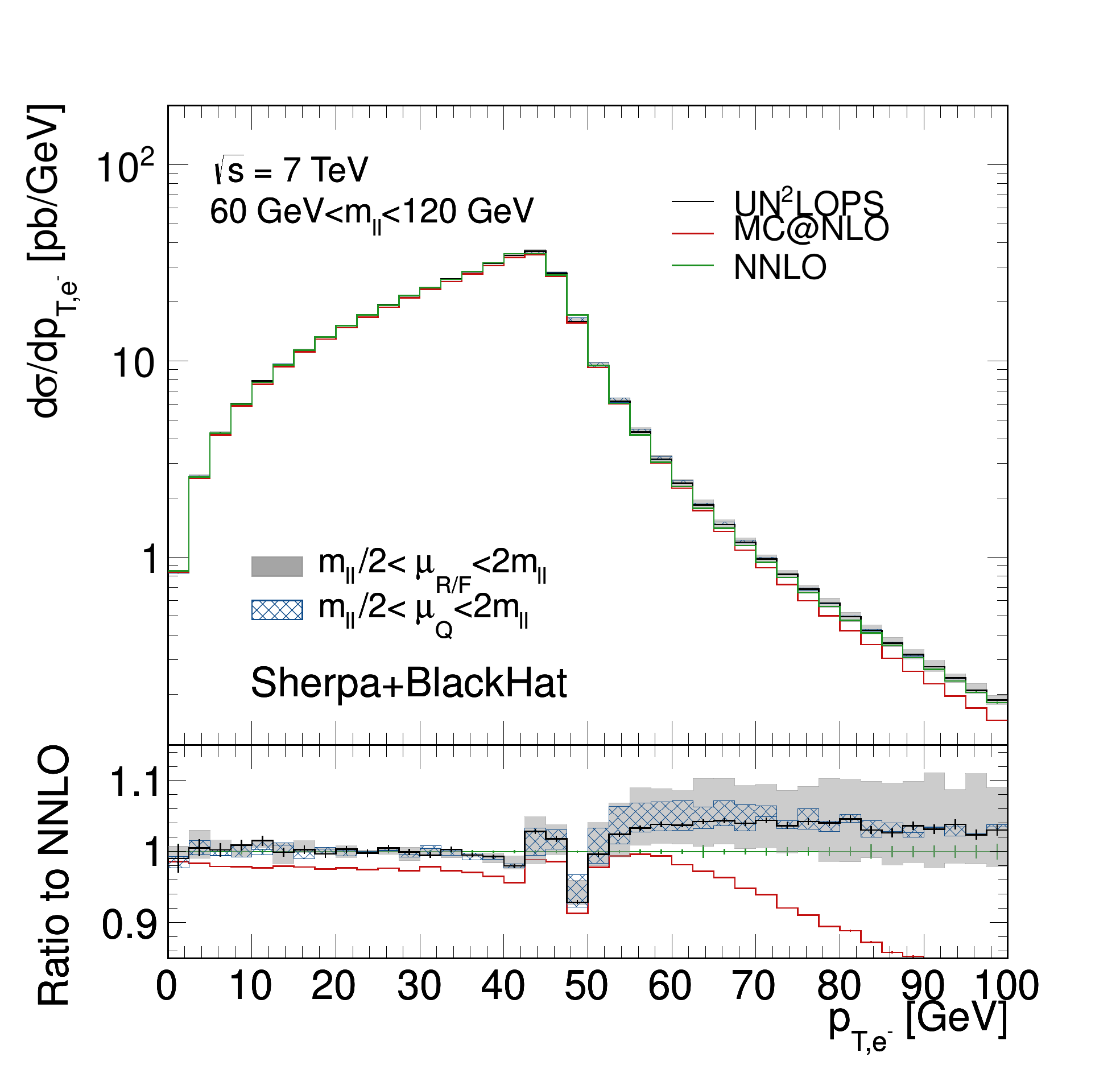}
  \caption{
    \protect\NNLOPS matched results for the Higgs boson \pT\ in gluon fusion with 
    \MINLO (left) and the \pT of the electron in Drell-Yan. 
    Figures taken from \cite{arXiv:1309.0017} and \cite{arXiv:1405.3607}.
    \label{fig:nnlops}
  }
\end{figure}

Recently, also NNLO computations have been matched to parton showers
The available implementations, however, are currently limited to singlet 
production where the logarithmic accuracy of the parton showers is 
sufficient for this task. 
Three unique formulations exist: \MINLO \cite{arXiv:1206.3572,arXiv:1309.0017,
  arXiv:1407.2940,arXiv:1603.01620,arXiv:1804.08141}, 
\UNtwoLOPS \cite{arXiv:1405.3607,arXiv:1407.3773} and 
\Geneva \cite{arXiv:1508.01475}, employing very different matching 
algorithms. 
Selected results are shown in Fig.\ \ref{fig:nnlops}.

\section{Multijet merging at next-to-leading order accuracy}

While \LOPS, \NLOPS and \NNLOPS describe observables dominated by 
topologies of a single jet multiplicity to the given accuracy, a large 
and important class of observables at LHC receives significant contributions 
from multiple jet multiplicities. 
Examples are $H_T$ and \pT spectra as well as azimuthal separations. 
Here, the accuracy of the above described calculations rapidly deteriorates. 
Multijet merging techniques were introduced to consistently combine 
calculations from various successive jet multiplicities at the highest 
available precision. 
At the same time, multijet merged event samples provide the LHC experiments 
with the largest freedom of projecting these samples onto as many observables 
as possible without the loss of accuracy. 
Available multijet merging methods, be it at LO or NLO accuracy, fall 
into two type of algorithms: CKKW-type and MLM-type.

\begin{figure}[t!]
  \begin{minipage}{0.45\textwidth}
    \lineskip-1.3pt
    \includegraphics[width=\textwidth]{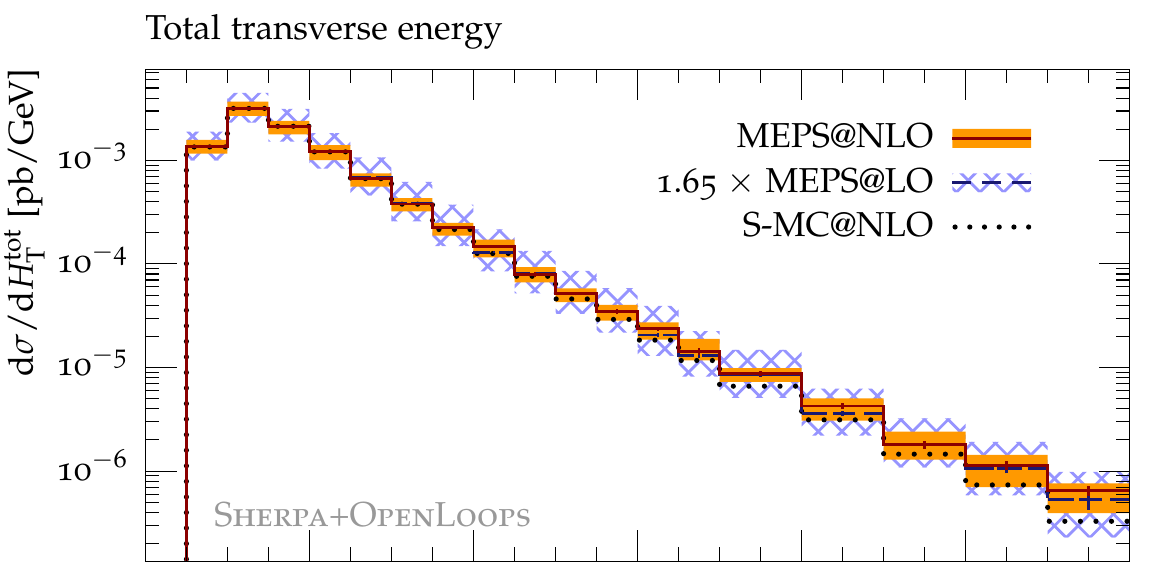}\\
    \includegraphics[width=\textwidth]{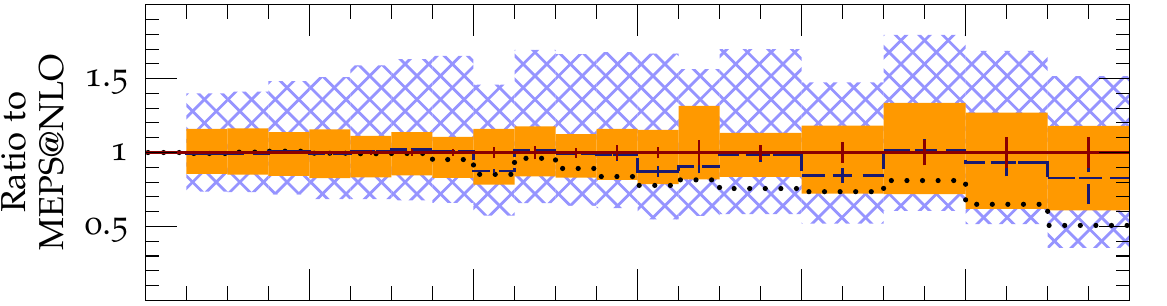}\\
    \includegraphics[width=\textwidth]{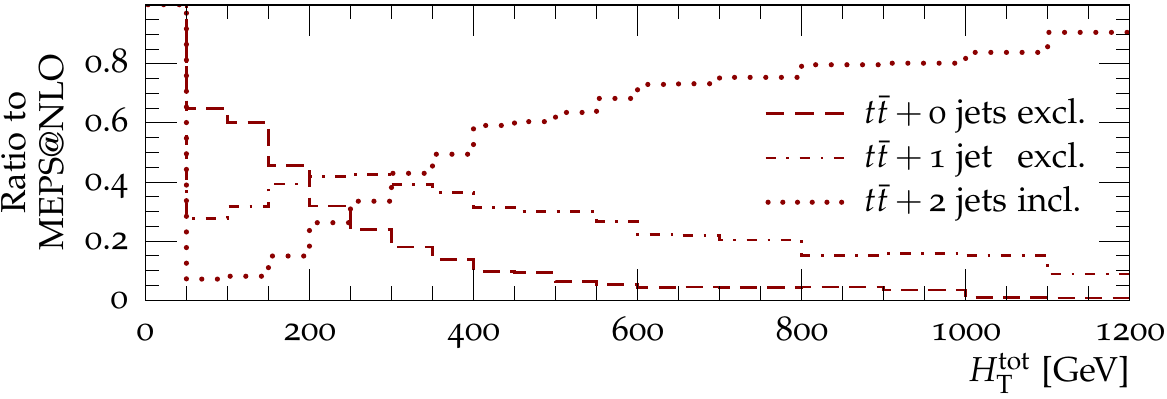}
  \end{minipage}
  \hfill
  \begin{minipage}{0.45\textwidth}
    \lineskip-1.3pt
    \includegraphics[width=\textwidth]{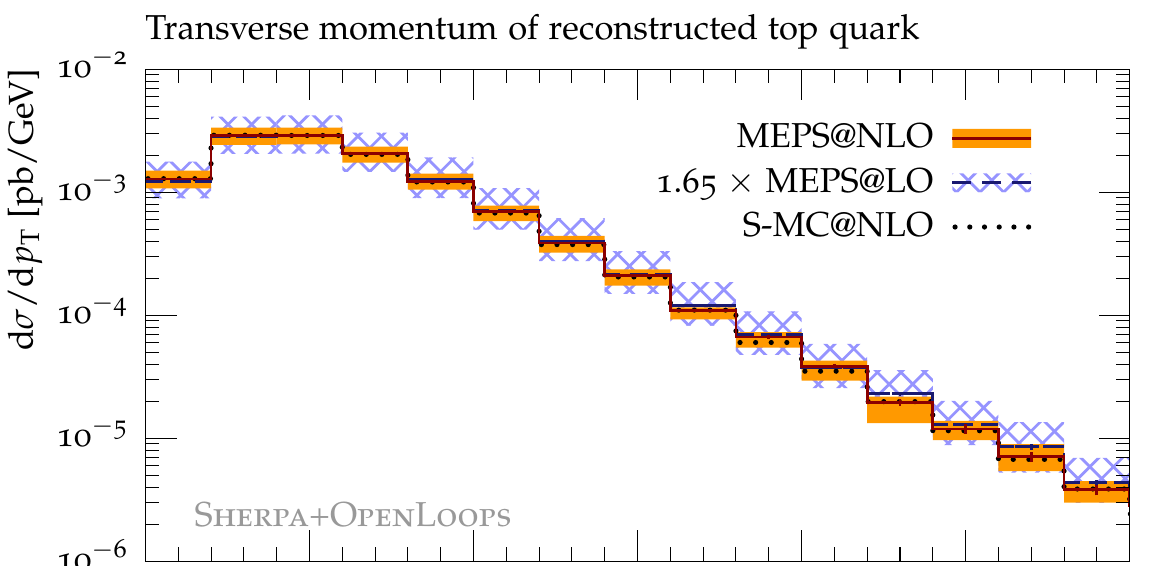}\\
    \includegraphics[width=\textwidth]{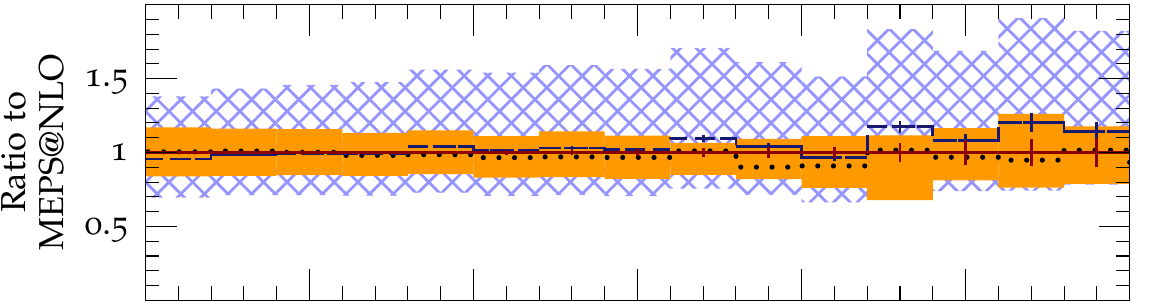}\\
    \includegraphics[width=\textwidth]{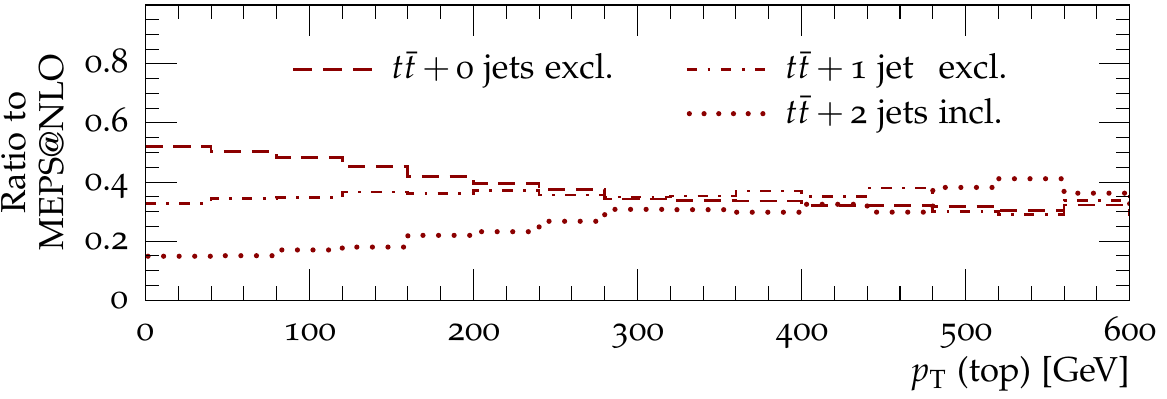}
  \end{minipage}
  \caption{
    Total transverse energy and top quark transverse momentum 
    deecribed with \protect\MEPS and \protect\MEPSatNLO with reduced theoretical 
    uncertainty.
    Figures taken from \cite{arXiv:1402.6293}.
    \label{fig:mepsnlo-ttbar}
  }
\end{figure}

\begin{figure}[t!]
  \includegraphics[width=0.48\textwidth]{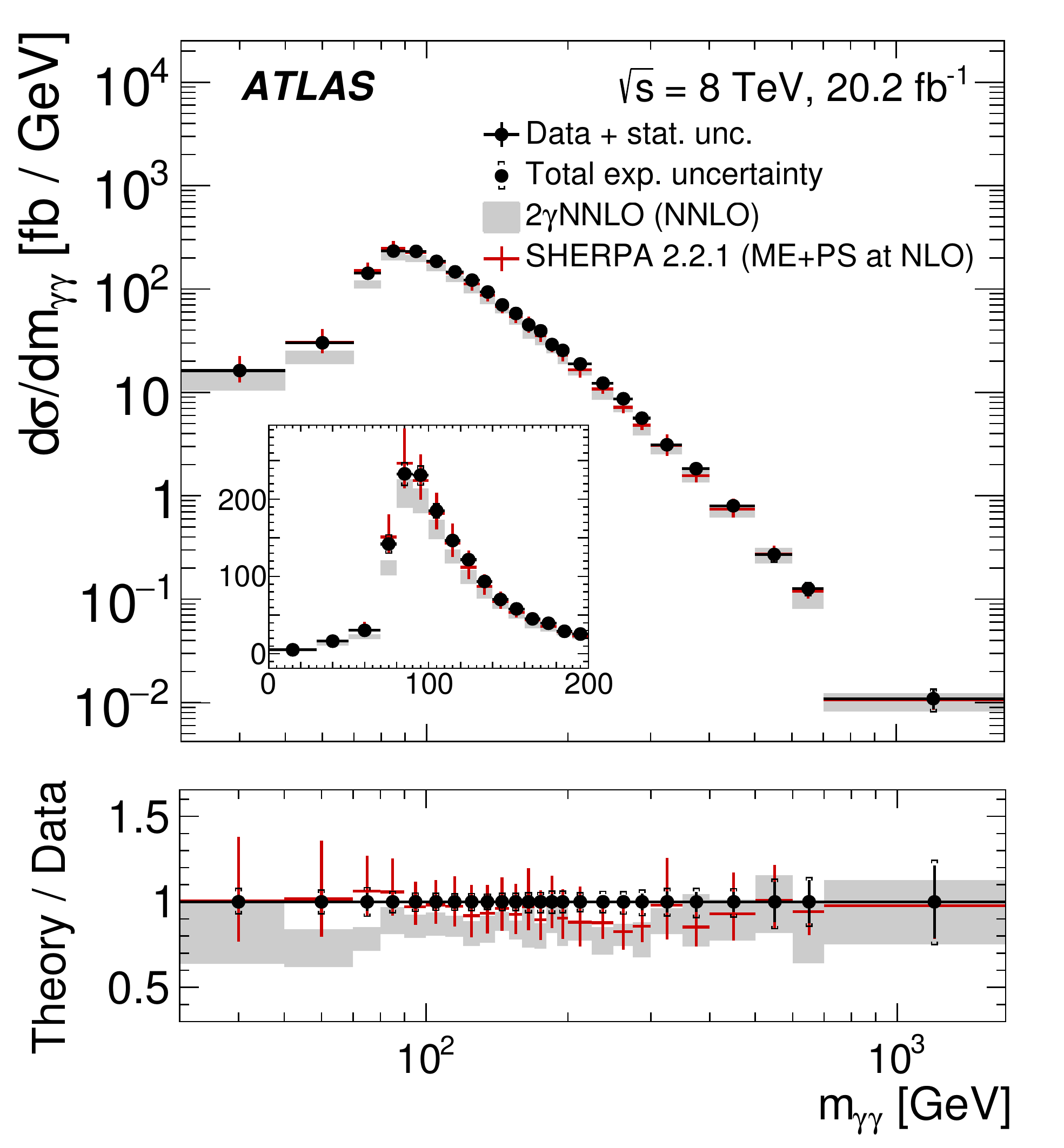}\hfill
  \includegraphics[width=0.48\textwidth]{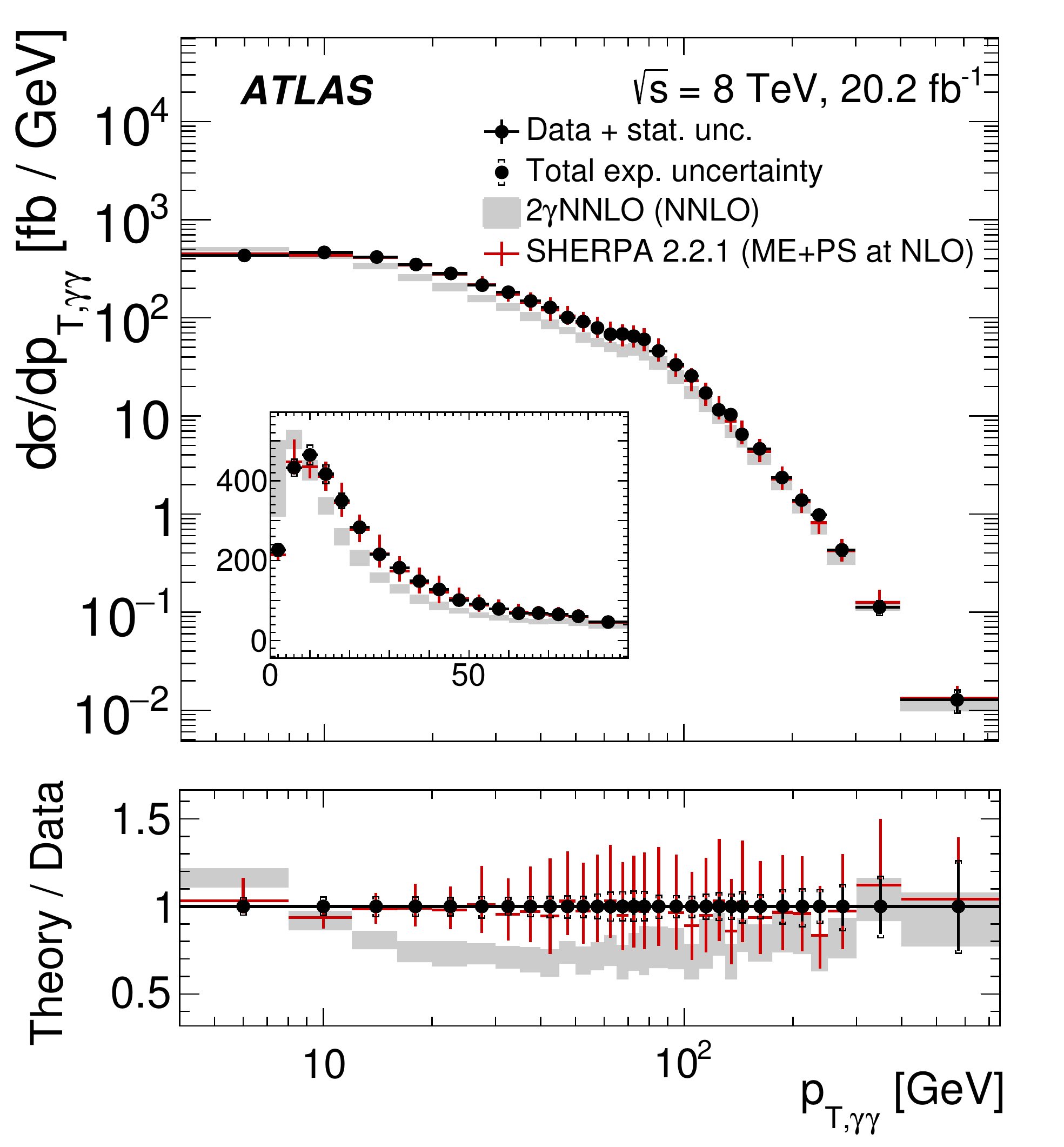}
  \caption{
    Diphoton invariant mass and pair transverse momentum in 
    photon pair production with \protect\MEPSatNLO compared 
    to ATLAS data.
    Figures taken from \cite{arXiv:1704.03839}.
    \label{fig:mepsnlo-yy}
  }
\end{figure}

Both algoithms define a resultion criterion \Qcut, which separates 
$n$-jet production from $(n+1)$-jet production. 
In this way the procedure can be iterated, adding jet multiplicities 
as long as it computationally feasible.
While the MLM-type algorithms merge either jet multiplicities described 
at LO (MLM \cite{hep-ph/0108069}) or NLO (FxFx \cite{arXiv:1209.6215}) only, 
the CKKW-type algorithms can merge successively either LO matrix elements 
to LO matrix elements
(\MEPS \cite{hep-ph/0109231,hep-ph/0112284,hep-ph/0205283,arXiv:0706.2569,
  arXiv:0903.1219,arXiv:1109.4829,arXiv:1211.4827,arXiv:1211.5467}), 
NLO matrix elements to LO matrix elements
(\MENLOPS \cite{arXiv:1004.1764,arXiv:1009.1127,arXiv:1207.5031,
  arXiv:1403.7516,arXiv:1803.00950}), 
or NLO matrix elements to NLO matrix elements
(\MEPSatNLO \cite{arXiv:1207.5030,arXiv:1207.5031,arXiv:1211.5467,
  arXiv:1211.7049,arXiv:1211.4827,arXiv:1211.7278,arXiv:1306.2703,
  arXiv:1401.7971,arXiv:1402.6293,arXiv:1611.07226,arXiv:1708.06283}).
Example results are shown in Figs.\ \ref{fig:mepsnlo-ttbar} and 
\ref{fig:mepsnlo-yy}.

Further, loop-induced process form an important class of 
processes measured at the LHC. 
Their theoretical description is complicated that the LO 
calculation already involves a loop computation. 
Nonetheless, additional jet activity is also prevalent in 
this class and a multijet merged calculation is desirable. 
Therefore, two different ansatzes have been persued. 
In cases where an effective theory which integrates out 
the loop exists, the calculation is performed in that 
effective theory (which has only LO complexity) and 
then reweighted to include the corrections due to the exact 
loop dynamics \cite{arXiv:1410.5806,arXiv:1604.03017} 
\nocite{arXiv:1506.01016,arXiv:1608.01195,arXiv:1611.00767}. 
In this way, a NLO merging in the effective theory including 
Born level loop corrections are feasible.

\begin{figure}[t!]
  \includegraphics[width=0.47\textwidth]{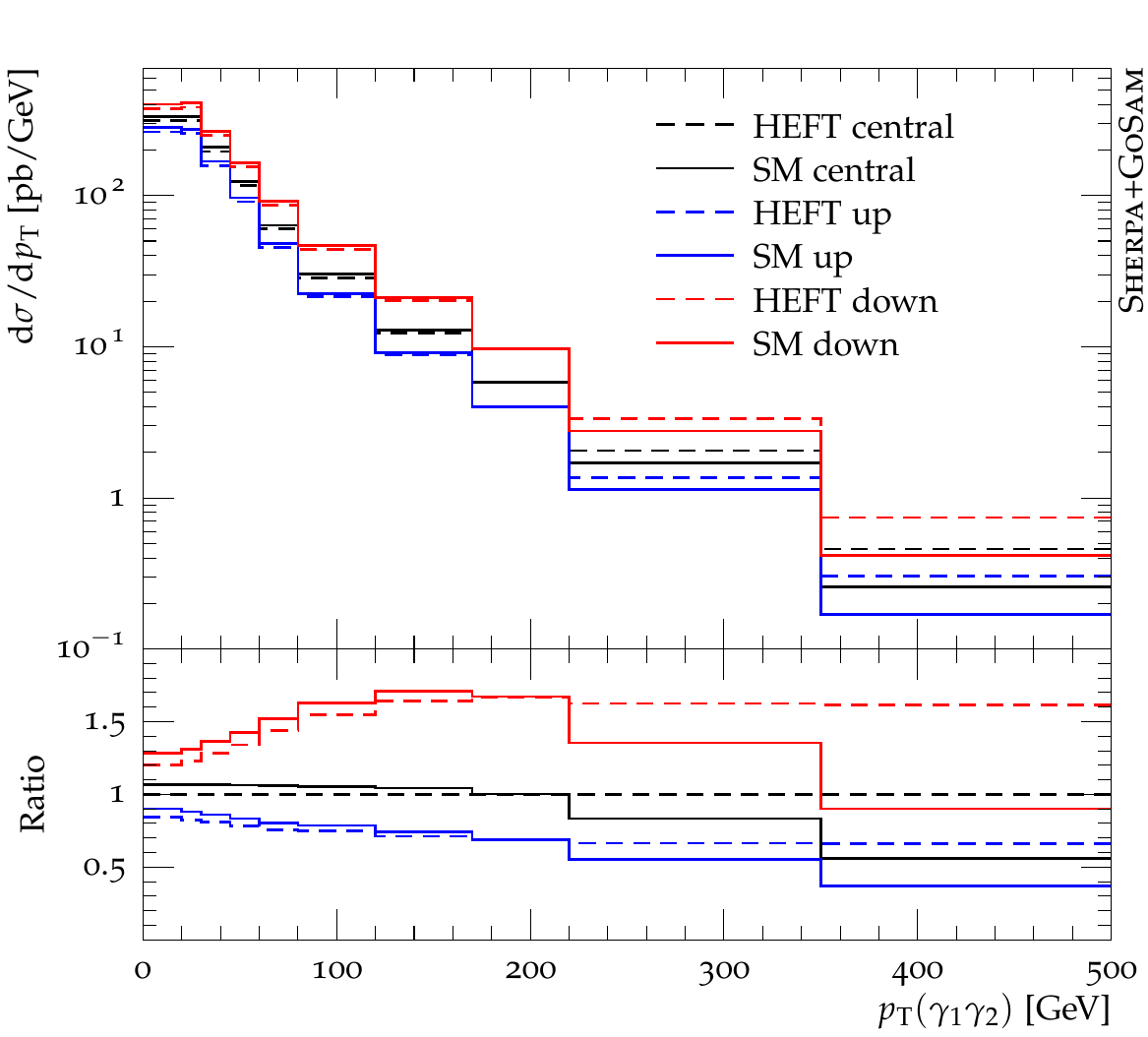}\hfill
  \includegraphics[width=0.47\textwidth]{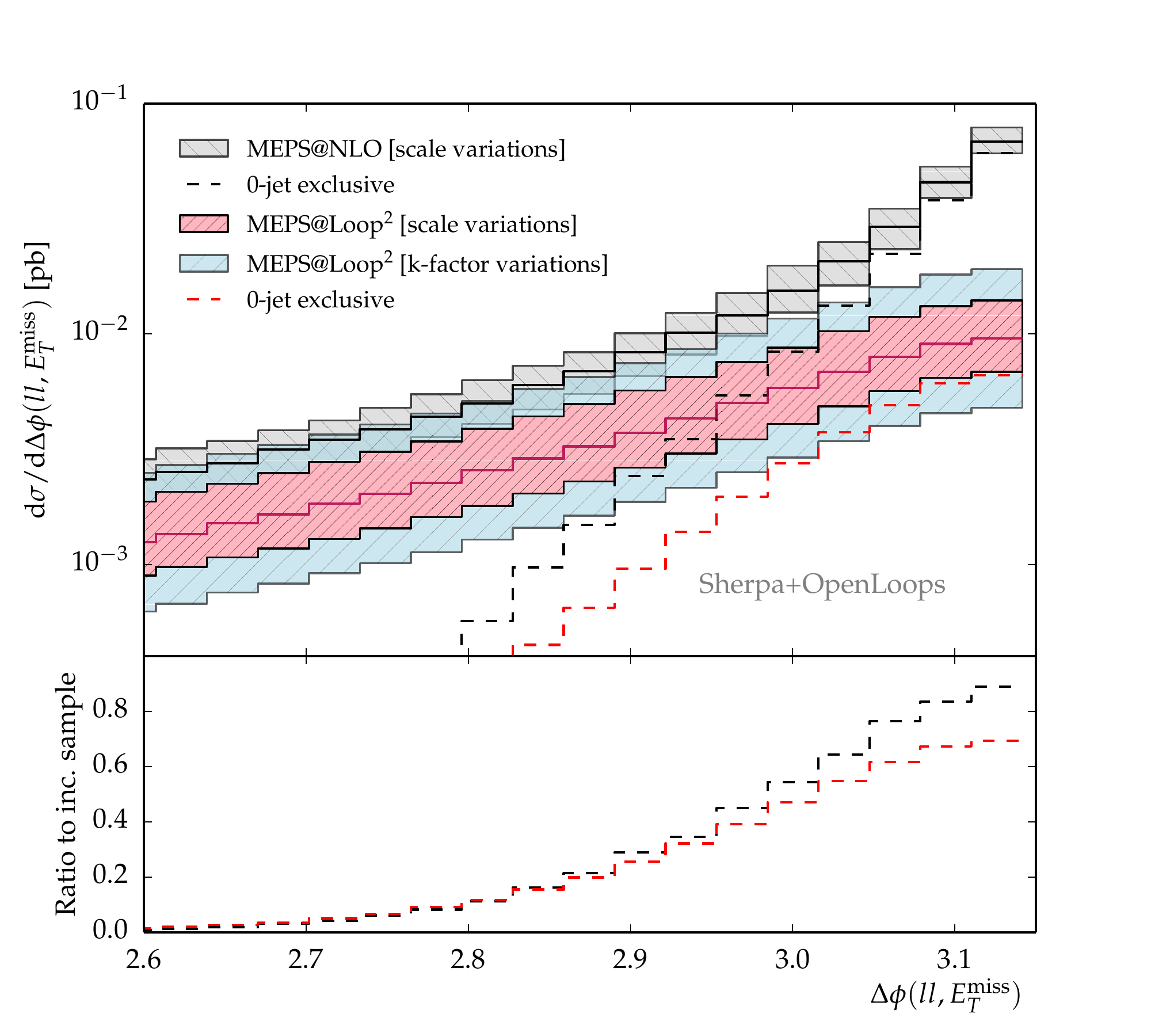}
  \caption{
    \textbf{Left:} 
    Top mass corrections in an NLO multijet merged calculation 
    of the diphoton \pT\ spectrum in Higgs production in gluon fusion.
    \textbf{Right:} 
    NLO $q\bar{q}$- and LO loop-induced $gg$-induced contribution 
    to $Zh$ associated production.
    Figure taken from \cite{arXiv:1509.01597}.
    \label{fig:mepsloop}
  }
\end{figure}

On the other hand, at LO accuracy, direct loop-induced 
calculations can be merged using a technique dubbed 
\MEPSatLoop \cite{arXiv:1309.0500,arXiv:1509.01597}. 
Fig.\ \ref{fig:mepsloop} shows example results.

Finally, approximate NLO EW corrections were incorporated in 
NLO multijet merging methods in \cite{arXiv:1511.08692}.

\section{Conclusions}

All processes of relevance to LHC analyses are available at least 
at \NLOPS accuracy. 
Key processes like $W$, $Z$ and $h$ production are even known to 
\NNLOPS accuracy. 
To also merge non-colour-singlet process at this accuracy to parton 
showers, the logarithmic accuracy of the latter need to be improved 
first \cite{arXiv:1611.00013,arXiv:1705.00982,arXiv:1705.00742,
  arXiv:1711.03497}.

On another, almost orthogonal direction, the recent progress in the 
automation of NLO EW corrections \cite{arXiv:1712.07975,arXiv:1412.5157,
  arXiv:1511.08692,arXiv:1606.02330,arXiv:1705.00598,arXiv:1705.04664,
  arXiv:1706.09022,arXiv:1710.11514,arXiv:1803.00950}
should be universally matched to parton showers and included in the 
standard event simulation. 
Current, approximate methods \cite{arXiv:1511.08692} are only targeted 
at large-\pT\ physics. 
At the same time, developments in including EW effects into 
parton showers \cite{arXiv:1401.5238,arXiv:1403.4788,arXiv:1611.00788,
  arXiv:1703.08562,arXiv:1712.07147} are not only relevant for 
future colliders at energies up to 100 TeV, but also offer a way 
of extending multijet merging to include jet emissions through 
hadronically decaying vector boson production as well.

\end{document}